\documentclass[conference]{IEEEtran}

\usepackage{graphicx}
\usepackage{amsmath,amssymb,amsfonts}
\usepackage{bm}
\usepackage{booktabs}
\usepackage{multirow}
\usepackage{algorithm}
\usepackage{algpseudocode}

\usepackage{pgfplots}
\pgfplotsset{compat=1.18}
\usepackage[caption=false,font=footnotesize]{subfig}
\usepackage{siunitx}
\usepackage{tikz}
\usepackage{float}
\usepackage{stfloats}
\usepackage{xcolor}
\usetikzlibrary{arrows.meta, positioning, shadows}

\makeatletter
\newcommand{\linebreakand}{%
  \end{@IEEEauthorhalign}
  \hfill\mbox{}\par
  \mbox{}\hfill\begin{@IEEEauthorhalign}
}
\makeatother

\usepackage{changes}

\usepackage[hidelinks]{hyperref}
\hypersetup{
    pdfauthor={Mahdi Taheri et al.},
    pdftitle={REQAP: Resilient Weight Packing and Quantization for DNN Acceleration}
}

\usepackage{fancyhdr}
\fancypagestyle{firstpage}
{
    \fancyhead[L]{\footnotesize © 2026 IEEE.  Personal use of this material is permitted.  Permission from IEEE must be obtained for all other uses, in any current or future media, including reprinting/republishing this material for advertising or promotional purposes, creating new collective works, for resale or redistribution to servers or lists, or reuse of any copyrighted component of this work in other works. This paper is accepted at the IEEE Baltic Electronics Conference 2026 (BEC2026).}
    \fancyhead[R]{}
}

\begin{document}

\title{REQAP: Resilient Weight Packing and Quantization for Edge DNN Acceleration}


\author{
\IEEEauthorblockN{
Mahdi Taheri$^{1,2}$, Samira Nazari$^{3}$, Mubassher Ansari$^{4}$, Ali Azarpeyvand$^{3}$
}
\linebreakand
\IEEEauthorblockN{
Mohsen Afsharchi$^{3}$, Maksim Jenihhin$^{2}$, and Christian Herglotz$^{4}$
}
\vspace{1em}
\IEEEauthorblockA{
$^{1}$Humboldt University of Berlin, Berlin, Germany\\
$^{2}$Tallinn University of Technology, Tallinn, Estonia\\
$^{3}$University of Zanjan, Zanjan, Iran\\
$^{4}$Brandenburg University of Technology Cottbus-Senftenberg, Cottbus, Germany
}
}

\maketitle
\thispagestyle{firstpage}
\begin{abstract}
Efficient deployment of Deep Neural Networks (DNNs) on edge accelerators
requires aggressive model compression while maintaining reliability in
fault-prone hardware environments. This paper presents a reliability-aware
quantized weight packing methodology for systolic-array-based DNN
accelerators. A sensitivity-driven mixed-precision quantization framework
assigns layer-wise bit-widths according to accuracy impact while enforcing
symmetric precision between weights and activations. A deterministic
register-level packing strategy consolidates multiple heterogeneous
operand pairs into fixed-width register words, enabling
SIMD-within-a-register (SWAR) style parallel execution that reduces both
memory footprint and execution cycles. To improve resilience against
hardware faults, selective bit-level protection replicates the most
significant bits (MSBs) of critical layers into unused register space,
achieving TMR-style protection with minimal overhead. A systolic-array
simulation framework is developed to evaluate the proposed packing and
fault-tolerance mechanisms under realistic execution conditions. Simulation in AlexNet, VGG-11, and ResNet-18 demonstrate up to 62\% memory reduction and up to 56\% reduction in Multiply-Accumulate (MAC) operations, while significantly improving accuracy resilience under fault injection compared to baseline and fully protected models.The proposed methodology achieves favorable trade-offs between reliability-aware performance (RAP) and accuracy degradation probability ($P_{drop}$) across varying bit-error rates, making it suitable for reliable edge inference.
\end{abstract}


\section{Introduction}

Quantization mitigates the extensive memory demand and challenge in Deep Neural Networks (DNNs) deployments by representing the network
parameters and activations using reduced numerical precision, thereby
decreasing memory footprint and computational complexity
\cite{taheri2024adam0,nazari2025reliability,jacob2018quantization,taheri2023deepaxe}.
Early quantization approaches employ homogeneous precision across all
network layers \cite{quant, fusion,nazarifortune}. However,
DNN layers exhibit varying sensitivity to quantization errors
\cite{vlsi-soc,taheri2024exploration,wang2019haq}, motivating
mixed-precision techniques that assign bit-widths according to the
importance of each layer \cite{jacob2018quantization,wang2019haq}.

In safety-critical domains such as autonomous driving and medical
diagnostics, reliable operation is equally important. DNN inference
executed on edge hardware is vulnerable to transient and permanent
faults arising from voltage scaling, radiation, or manufacturing
variability \cite{eslami2024mono}. Hardware-level
fault tolerance is therefore frequently introduced using
Error-Correcting Codes (ECC), redundancy, or bit-level protection
mechanisms \cite{nazarifortune}. While these approaches
improve resilience, they often incur significant overhead in memory,
area, and energy.

Existing mixed-precision quantization methods are primarily designed
to preserve model accuracy and typically rely on retraining
\cite{jacob2018quantization} or computationally expensive neural
architecture search techniques \cite{wang2019haq}. In addition, fault
protection mechanisms are commonly applied independently of quantization
decisions \cite{eslami2024mono,nazarifortune}. As a result, opportunities
to jointly optimize quantization, execution efficiency, and reliability
remain largely unexplored.

At the hardware level, register-level parallelism provides an additional
opportunity for improving efficiency. SIMD-Within-A-Register (SWAR)
techniques exploit subword-level parallelism by packing multiple low-bit
operands into fixed-width registers~\cite{sharma2018bit}. Recent work
applies such concepts to DNN inference, including ULPPACK
\cite{won2023ulppack} for sub-8-bit matrix multiplication and Bit Fusion
\cite{sharma2018bit} for dynamic bit-width composition in systolic
arrays. However, existing approaches generally assume homogeneous
operand widths or rely on runtime hardware reconfiguration. Moreover,
they do not integrate sensitivity-driven quantization or fault
protection within the packing process.

Consequently, current approaches treat quantization, register packing,
and reliability as largely independent optimization problems. This
separation prevents exploitation of the natural register slack created
by mixed-precision representations and limits the ability to embed
low-overhead fault protection directly within packed data structures.

To address this limitation, an integrated methodology is introduced that
co-optimizes mixed-precision quantization, deterministic register
packing, and selective fault protection for systolic-array-based DNN
accelerators. A systolic-array simulation framework is developed to
evaluate the execution efficiency and reliability characteristics of
the proposed approach under realistic fault conditions.

The main contributions of this paper are as follows:

\begin{itemize}

\item \textbf{Co-Designed Mixed-Precision Quantization and Register Packing:}
A methodology that jointly determines layer-wise quantization precision
and deterministic register packing to improve execution efficiency,
memory utilization, and inference reliability.

\item \textbf{Selective Bit-Level Protection via Register Slack:}
A fault-tolerance mechanism that replicates the most significant bits
of sensitive layers into unused register space, enabling TMR-style
fault masking with minimal memory overhead.

\item \textbf{Deterministic SWAR Packing Algorithm:}
A Safe First-Fit Decreasing (Safe-FFD) packing strategy that packs
heterogeneous-precision operand pairs into fixed-width registers at
compile time, enabling efficient SIMD-style execution without hardware
modification.

\item \textbf{Systolic-Array Evaluation Framework:}
A systolic-array-based simulation environment developed to model
register-level packing, execution behavior, and fault injection for
reliable DNN inference evaluation.

\end{itemize}
The remainder of the paper is organized as follows.
Section~\ref{sec:methodology} describes the proposed methodology.
Section~\ref{sec:results} presents the experimental setup and evaluation
results. Section~\ref{sec:conclusion} concludes the paper.

\section{Proposed Methodology}
\label{sec:methodology}
The methodology introduced in this paper combines Single Instruction, Multiple Data (SIMD) register-level packing with selective bit-level redundancy to balance performance, memory efficiency, and fault resilience.

\subsection{Packing and Quantization}

The quantization framework assigns bit-widths to each layer through 
sensitivity analysis. Each layer is individually quantized at candidate 
bit-widths, and the accuracy drop relative to full precision is recorded. A greedy search reduces precision from the least sensitive layers until a target accuracy threshold $\tau$ is reached, producing a bit-width profile $\mathbf{b} = [b_1, b_2, \ldots, b_K]$, where $b_i$ is the bit-width assigned to operand slot $i$ and $K$ is the total number of slots across all layers of the network.

To enable efficient register packing, a symmetric precision constraint is enforced: the bit-width of each weight must equal that of its corresponding activation, i.e., \(\mathrm{bits}(W_i) = \mathrm{bits}(A_i) = b_i\), ensuring deterministic packing.

Once the bit-widths are assigned, operand pairs are packed into fixed-width register words using a \textit{Safe First-Fit Decreasing (Safe-FFD)} strategy. Each pair requires a slot cost $\phi_i = 2b_i$, representing the combined storage of its weight and activation. The slots are sorted in non-increasing order of $\phi_i$ and inserted into the first existing word that satisfies both the capacity and overflow constraints. If no word fits, a new word is created. No operand crosses a register boundary.

The packing constraint for register word $B_j$ is:
\begin{equation}
\sum_{i \in B_j} 2b_i \leq R
\end{equation}
where $R$ is the register width and $B_j$ is the set of slots in word $j$.

The packing density $d_j$ is defined as the number of operand pairs 
successfully packed into word $j$. The average packing density across 
all packed words is defined as:
\begin{equation}
    \bar{d} = \frac{1}{|\mathcal{J}|} \sum_{j} d_j
\end{equation}

Any unused bits within a word constitute slack, denoted as:
\begin{equation}
    s_j = R - \sum_{i \in B_j} 2b_i
\end{equation}

To prevent overflow during fixed-point accumulation, an additional constraint is enforced:
\begin{equation}
    \sum_{i \in B_j} (2^{b_i} - 1)^2 < 2^R
\end{equation}

This overflow constraint ensures that the accumulated intermediate values remain within the representable range of the register.
\begin{algorithm}[h]
\caption{Safe First-Fit Decreasing (Safe-FFD) Register Packing}
\label{alg:safe_ffd}
\begin{algorithmic}[1]
\Require Bit-width profile $\mathbf{b} = [b_1,b_2,\ldots,b_K]$, register width $R$
\Ensure Packed word sequence $\mathcal{J}$, density list $\mathcal{D}$
\State Compute slot costs $\phi_i \leftarrow 2b_i$ for all slots $i$
\State Sort all slots by $\phi_i$ in non-increasing order to obtain sorted list $\mathcal{S}_{\text{slot}}$
\State Initialize $\mathcal{J} \leftarrow \emptyset$, $\mathcal{D} \leftarrow \emptyset$, $\textit{bins} \leftarrow \emptyset$
\For{each slot $i \in \mathcal{S}_{\text{slot}}$}
    \State $\textit{placed} \leftarrow \textbf{false}$
    \For{each existing bin $B_j \in \textit{bins}$}
        \If{\textsc{StorageFit}$(B_j,\phi_i,R)$ \textbf{and} \textsc{OverflowSafe}$(B_j,i,R)$}
            \State Insert slot $i$ into $B_j$
            \State $\textit{placed} \leftarrow \textbf{true}$
            \State \textbf{break}
        \EndIf
    \EndFor
    \If{\textbf{not} $\textit{placed}$}
        \State Open new bin $B_{\text{new}} \leftarrow \{i\}$
        \State Append $B_{\text{new}}$ to \textit{bins}
    \EndIf
\EndFor
\For{each bin $B_j \in \textit{bins}$}
    \State Append $B_j$ to $\mathcal{J}$
    \State Append $|B_j|$ to $\mathcal{D}$
\EndFor
\State \textbf{return} $\mathcal{J}, \mathcal{D}$
\end{algorithmic}
\vspace{2pt}\hfill\includegraphics[height=1.8em]{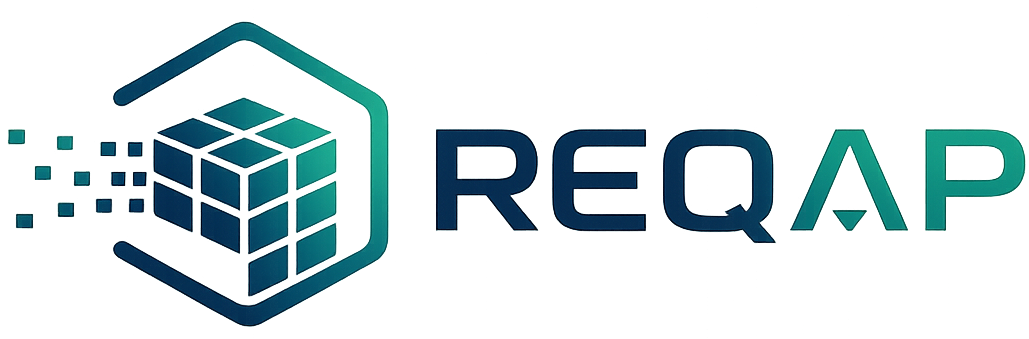}\vspace{-2pt}
\end{algorithm}

Algorithm~\ref{alg:safe_ffd} formalizes Safe-FFD packing. Slots are sorted by cost (Line 2), then iteratively placed into the first feasible bin (Lines 4--13). If no bin fits, a new bin opens (Line 12). The algorithm returns packed words and densities (Lines 15--17).

The storage and overflow safety predicates used in Algorithm~\ref{alg:safe_ffd} are defined as:
\begin{equation}
\textsc{StorageFit}(B_j,\phi_i,R):
\quad
\sum_{k \in B_j}\phi_k + \phi_i \leq R,
\end{equation}
\begin{equation}
\textsc{OverflowSafe}(B_j,i,R):
\quad
\sum_{k \in B_j \cup \{i\}} (2^{b_k}-1)^2 < 2^R.
\end{equation}

Packing reduces iteration depth from $K$ slots to $K_{\mathrm{eff}} = |\mathcal{J}|$ words, lowering memory accesses and cycles proportionally to $\bar{d}$.

\subsection{Sensitivity-Based Layer Profiling}

To balance accuracy preservation and packing density, the framework first profiles each layer’s sensitivity to quantization. The sensitivity of layer $l$ at bit-width $b \in \mathcal{B}$ is defined as the accuracy drop relative to the full-precision reference:
\begin{equation}
S_{l,b} = \text{Acc}_{\text{REF}} - \text{Acc}_{\text{QUANT}}(l,b \mid \mathbf{b}_{\text{others}} = b_{\text{max}}) \end{equation} where $\mathcal{B} = \{b_{\text{min}}, \ldots, b_{\text{max}}\}$ is the candidate bit-width set, $\text{Acc}_{\text{REF}}$ is the full-precision accuracy, and $\text{Acc}_{\text{QUANT}}$ is the accuracy when only layer $l$ is quantized to $b$ bits.

Using this profile, we apply a greedy search to assign the bit-width vector $\mathbf{b} = [b_1, b_2, \dots, b_L]$ under the accuracy-drop constraint $\tau$:
\begin{equation}
\arg \min_{\mathbf{b}} \sum_{l=1}^{L} \text{Bins}(\text{Safe-FFD}(b_l, R))
\quad \text{s.t.} \quad
\sum_{l=1}^{L} S_{l,b_l} \leq \tau
\end{equation}
where $\text{Bins}(\text{Safe-FFD}(b_l, R))$ is the number of register words required for layer $l$ under register width $R$, computed with Safe First-Fit Decreasing (Algorithm~1).

This importance-driven profiling ensures that critical layers retain higher precision, while resilient layers are aggressively compressed to maximize packing density $d_j$. The resulting bit-width schedule balances inference accuracy with hardware execution efficiency.

\subsection{Selective Protection with MSB Redundancy}

To improve reliability without uniform overhead, the most critical layers are selectively protected. For these layers, the most significant bit (MSB) of each quantized weight is replicated twice into available empty bits within the register word, enabling Triple Modular Redundancy (TMR). These replicated bits do not alter the quantization range or bit-width; they are used solely for fault masking.

During inference, a lightweight majority voter reconstructs the MSB based on the three replicas. The redundant bits are then logically cleared, and the packed operands proceed to multiplication. This mechanism enhances fault resilience with minimal area and memory overhead while maintaining the packing efficiency established during quantization.

\subsection{Systolic-Array Simulation Framework for Packed SIMD Execution}

To evaluate the proposed co-designed quantization, packing, and protection strategy under realistic execution conditions, a systolic-array simulation framework targeting FPGA-style accelerators is developed. The framework models register-level SIMD execution, packed data movement and accumulation behavior.

The simulator instantiates a parameterized systolic array in which each Processing Element (PE) operates on fixed-width registers (e.g., $R=16$ or $R=32$ bits) and consumes one packed operand word per cycle. Packed register words are generated offline using the Safe-FFD algorithm and provided to the simulator as execution traces. Each packed word may contain multiple heterogeneous operand pairs arranged according to the compile-time bit-width schedule.

Within each simulated PE, subword-level parallelism is modeled using statically determined bit offsets and masks. The SIMD Multiply-Accumulate (MAC) operation for packed word $j$ is expressed as:
\begin{equation}
P_j = \sum_{k=1}^{d_j} 
\left[ \left( W_j \gg \text{pos}_k \right) \,\&\, \text{mask}_k \right] 
\times 
\left[ \left( A_j \gg \text{pos}_k \right) \,\&\, \text{mask}_k \right],
\label{eq:mac}
\end{equation}
where $W_j$ and $A_j$ denote the packed weight and activation register words, respectively, and $d_j$ is the packing density of word $j$.

The compile-time offsets and masks are defined as:
\begin{equation}
\text{pos}_k = \sum_{i=1}^{k-1} 2b_i, 
\quad 
\text{mask}_k = (2^{2b_k} - 1),
\end{equation}
where $b_k$ represents the bit-width of the $k$-th operand pair within the packed word. The operators $\gg$ and $\&$ denote bitwise shift and AND operations, respectively.

Let $C_t$ denote the accumulated value in the PE's accumulator register 
at time step $t$. Accumulation proceeds as:
\begin{equation}
C_{t+1} = C_t + P_j,
\end{equation}
and the final dot-product output is computed as:
\begin{equation}
C = \sum_{j=1}^{K_{\text{eff}}} P_j,
\end{equation}
where $K_{\text{eff}}$ is the effective number of packed register words 
after Safe-FFD packing, reduced from the original $K$ operand pairs.

This execution model reduces iteration depth from $K$ to $K_{\text{eff}}$, lowering memory accesses and execution cycles proportionally to the average packing density $\bar{d}$. For protected layers, MSB replicas embedded in register slack $s_j$ are reconstructed via majority voting before computation. This correction operates in parallel with data movement, introducing negligible latency.

\begin{figure}[h]
    \centering
    \includegraphics[width=\columnwidth]{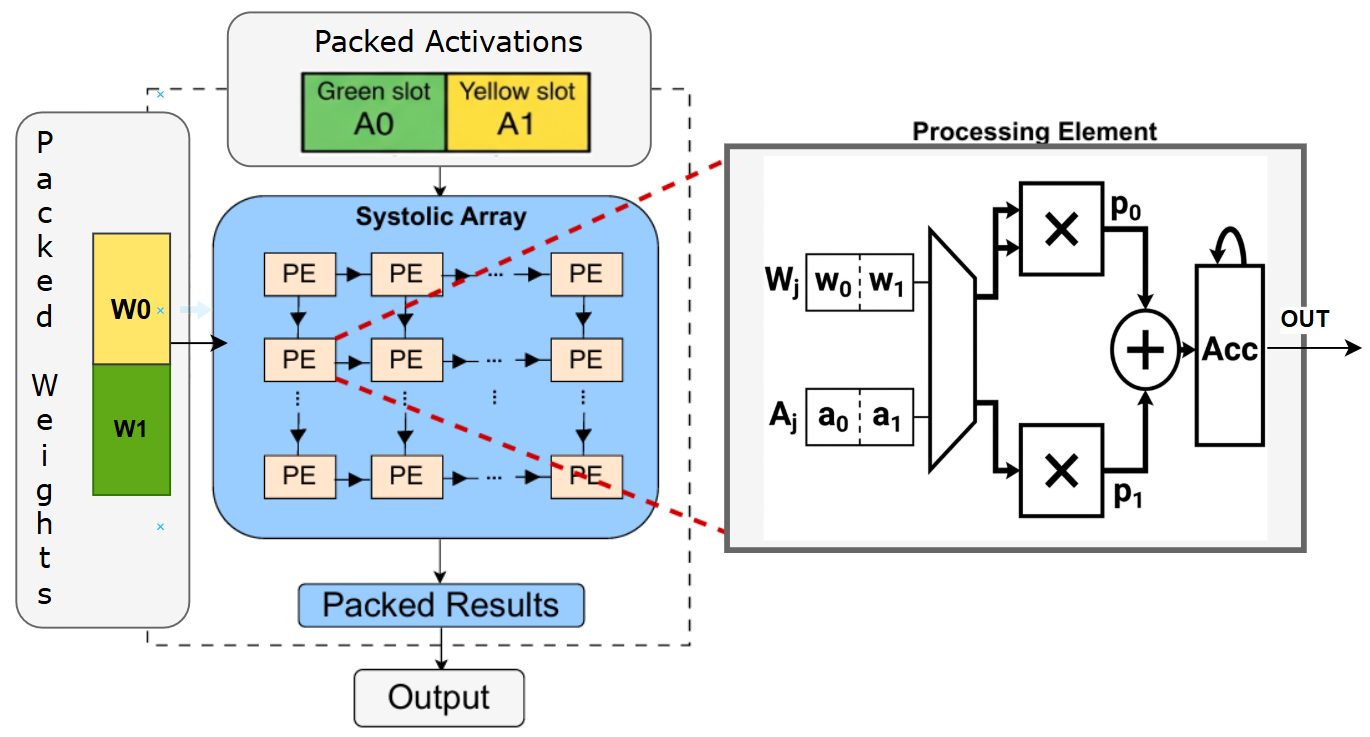}
    \caption{A high-level flow of methodology}
    \label{fig:hw_mapping}
\end{figure}

Figure~\ref{fig:hw_mapping} illustrates the complete methodology flow. Packed operands enter the systolic array, where each PE consumes one packed word per cycle and performs parallel multiply-accumulate operations. The majority voter corrects MSBs for protected layers, ensuring fault-resilient execution without modifying the underlying PE architecture.

\section{Experimental Results}
\label{sec:results}

\subsection{Experimental Setup}
The reliability and performance of the proposed method are evaluated using three neural networks: AlexNet, VGG-11, and ResNet-18. AlexNet is trained on the Fashion MNIST dataset, whereas VGG-11 and ResNet-18 are trained on the CIFAR-10 dataset. Each model undergoes sensitivity-based quantization, deterministic register packing, and reliability analysis, resulting in multiple configurations that balance computational efficiency, memory footprint, and fault resilience. 

To comprehensively assess the impact of the proposed methodology, we evaluate both \textbf{execution efficiency} and \textbf{reliability}. For execution efficiency, we measure the number of Multiply-Accumulate (MAC) operations, memory utilization, and packing density achieved through the Safe-FFD algorithm (Algorithm 1). For reliability, we employ two established metrics: \textbf{Reliability-Aware Performance (RAP)} and \textbf{Probability of Accuracy Drop ($P_{drop}$)} \cite{nazarifortune,nazari2025genie}.  

The RAP metric jointly captures the trade-off between accuracy degradation, memory overhead, and execution cost:
\begin{equation}
\label{eq:rap}
\text{RAP} = \Delta\text{acc} \times \text{mem\_ovh} \times 
\text{perf\_ovh}
\end{equation}
where $\text{mem\_ovh}$ and $\text{perf\_ovh}$ denote memory and execution overhead relative to the unprotected baseline, respectively~\cite{nazari2025genie}. A lower RAP reflects a better reliability-efficiency trade-off.

The $P_{drop}$ metric estimates the likelihood of an accuracy drop over the device's operational lifetime due to faults. This metric builds upon the probability of single-bit flips and serves as a comprehensive measure of the network's resilience to fault-induced errors.

\subsection{Reliability and performance evaluation}

Reliability is evaluated using random fault injections across all weights with BERs between $10^{-5}$ and $3\times10^{-4}$, covering a realistic range of hardware error rates. Each experiment is repeated sufficiently to ensure statistically reliable results following \cite{leveugle2009statistical}.

For each DNN model (AlexNet, VGG-11, and ResNet-18), representative configurations are selected at 3-, 4-, and 5-bit quantization levels, using the maximum feasible packing depth for a 32-bit register. Each packed configuration is compared with its unprotected counterpart and with a fully protected baseline using the same MSB replication and majority voting method as in \cite{nazarifortune}. Table~\ref{tab:config_labels} summarizes the notation used throughout the results.
\begin{table}[h]
\centering
\footnotesize
\caption{Notation Used for Model Configurations}
\label{tab:config_labels}

\begin{tabular}{c p{0.6\columnwidth}} 
\toprule
\textbf{Label} & \textbf{Description} \\ 
\midrule
$q3$ & 3-bit quantized model without protection \\
$qp3$ & 3-bit quantized model with protection applied to all layers \\
$mq$ & multi-bit quantized model without protection applied to all layers \\
$(p3,d4)$ & 3-bit quantized model, protected only at critical layers, packed with depth 4 \\
$q4$, $qp4$, $(p4,d3)$ & Same as above, using 4-bit quantization \\
$q5$, $qp5$, $(p5,d2)$ & Same as above, using 5-bit quantization \\
\bottomrule
\end{tabular}
\end{table}

\begin{table}[h]
\centering
\scriptsize
\setlength{\tabcolsep}{3pt}
\caption{Memory utilization, accuracy drop under fault injection, $P_{drop}$, and RAP at BER = $1.00 \times 10^{-4}$.}
\label{tab:all_data_single_ber}

\resizebox{\columnwidth}{!}{
\begin{tabular}{c c c c c c}
\toprule
\textbf{Model} & \textbf{Type} & \textbf{Memory} & \textbf{Acc. Drop (\%)} & \textbf{$P_{drop}$} & \textbf{RAP} \\
\midrule

\multirow{9}{*}{VGG-11}
& q5 & 140665280 & 30.77 & 8.10E-2 & 51.28 \\
& qp5 & 196931392 & 0.01 & 5.16E-5 & 8.97 \\
& (p5,d2) & 140898112 & 1.46 & 3.85E-3 & 5.15 \\
& q4 & 112532224 & 78.06 & 1.31E-1 & 113.46 \\
& qp4 & 168798336 & 0 & 0 & 0 \\
& (p4,d3) & 112765056 & 2.83 & 4.79E-3 & 9.18 \\
& q3 & 84399168 & 79.80 & 7.56E-2 & 79.80 \\
& qp3 & 140665280 & 0.07 & 1.84E-4 & 43.16 \\
& (p3,d4) & 84632000 & 9.06 & 8.63E-3 & 32.59 \\

\midrule

\multirow{9}{*}{ResNet-18}
& q5 & 55821760 & 68.71 & 2.85E-2 & 104.22 \\
& qp5 & 78150464 & 0.80 & 6.50E-4 & 195.30 \\
& (p5,d2) & 60781376 & 5.38 & 2.64E-3 & 2.12 \\
& q4 & 44657408 & 66.85 & 1.77E-2 & 109.09 \\
& qp4 & 66986112 & 0.14 & 8.36E-5 & 29.91 \\
& (p4,d3) & 49617024 & 8.25 & 2.70E-3 & 1.79 \\
& q3 & 33493056 & 64.58 & 9.64E-3 & 64.58 \\
& qp3 & 55821760 & 0.18 & 7.46E-5 & 30.21 \\
& (p3,d4) & 38452672 & 24.30 & 4.78E-3 & 3.48 \\

\midrule

\multirow{9}{*}{AlexNet}
& q5 & 291447520 & 6.46 & 7.30E-2 & 9.23 \\
& qp5 & 408026528 & 2.52 & 5.58E-2 & 1612.18 \\
& (p5,d2) & 291552672 & 0.51 & 5.77E-3 & 0.15 \\
& q4 & 233158016 & 16.13 & 1.17E-1 & 18.36 \\
& qp4 & 349737024 & 1.99 & 3.24E-2 & 1156.05 \\
& (p4,d3) & 233263168 & 3.14 & 2.27E-2 & 0.52 \\
& q3 & 174868504 & 43.31 & 1.76E-1 & 43.31 \\
& qp3 & 291447520 & 0.05 & 5.65E-4 & 34.05 \\
& (p3,d4) & 174973656 & 11.01 & 4.48E-2 & 0.76 \\

\bottomrule
\end{tabular}
}
\end{table}

Table~\ref{tab:all_data_single_ber} shows that selective protection combined with register-level packing provides a better trade-off than both unprotected and fully protected baselines. Across AlexNet, VGG-11, and ResNet-18, the proposed $(p,d)$ configurations reduce memory usage and execution cost while maintaining lower RAP and $P_{drop}$ values. In particular, higher packing depth improves efficiency, whereas selective MSB protection preserves fault tolerance without the overhead of full protection.

Notably, the $q3$ configurations show the worst fault tolerance and highest RAP; since $q3$ is the unprotected baseline ($mem\_ovh=perf\_ovh=1$ in Eq.~\eqref{eq:rap}), RAP reduces to $\Delta acc$, matching Table~\ref{tab:all_data_single_ber}, whereas protected/packed configurations have overhead terms deviating from unity. The $qp4$ case for VGG-11 similarly shows complete fault masking at this BER, as majority voting corrects every corrupted MSB, yielding zero accuracy drop and, since RAP and $P_{drop}$ are proportional to $\Delta acc$, zero for both metrics as well. By protecting only the most vulnerable layers with optimal packing, $(p3,d4)$ substantially improves resilience; e.g., in VGG-11 accuracy drop falls from 79.8\% ($q3$) to 9.06\% $(p3,d4)$.

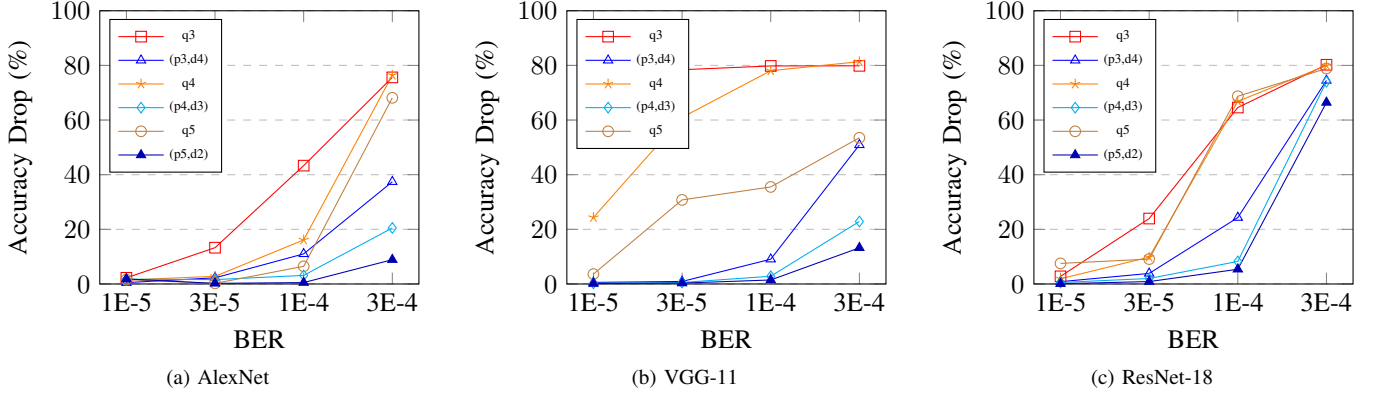
\begin{figure*}[ht]
    \centering
    \subfloat[AlexNet]{%
\begin{tikzpicture}
\begin{axis}[
    width=0.32\textwidth,
    height=0.22\textheight,
    xlabel={BER},
    ylabel={Accuracy Drop (\%)},
    ymin=0, ymax=100,
    xtick={1,2,3,4},
    xticklabels={1E-5, 3E-5, 1E-4, 3E-4},
    legend pos=north west,
    ymajorgrids=true,
    grid style=dashed,
    legend style={font=\tiny},
]
\addplot[color=red,mark=square] coordinates {(1,2.24)(2,13.28)(3,43.31)(4,75.61)};
\addlegendentry{q3}
\addplot[color=blue,mark=triangle] coordinates {(1,0.49)(2,2.25)(3,11.01)(4,37.41)};
\addlegendentry{(p3,d4)}
\addplot[color=orange,mark=star] coordinates {(1,1.45)(2,2.8)(3,16.13)(4,76.44)};
\addlegendentry{q4}
\addplot[color=cyan,mark=diamond] coordinates {(1,1.74)(2,1.6)(3,3.14)(4,20.5)};
\addlegendentry{(p4,d3)}
\addplot[color=brown,mark=o] coordinates {(1,1.59)(2,0.22)(3,6.46)(4,68.14)};
\addlegendentry{q5}
\addplot[color=blue!70!black,mark=triangle*] coordinates {(1,1.84)(2,0.22)(3,0.51)(4,8.89)};
\addlegendentry{(p5,d2)}
\end{axis}
\end{tikzpicture}
\label{fig:alex-rel}
}
\hfill
\subfloat[VGG-11]{%
\begin{tikzpicture}
\begin{axis}[
    width=0.32\textwidth,
    height=0.22\textheight,
    xlabel={BER},
    ylabel={Accuracy Drop (\%)},
    ymin=0, ymax=100,
    xtick={1,2,3,4},
    xticklabels={1E-5, 3E-5, 1E-4, 3E-4},
    legend pos=north west,
    ymajorgrids=true,
    grid style=dashed,
    legend style={font=\tiny},
]
\addplot[color=red,mark=square] coordinates {(1,72.51)(2,78.42)(3,79.80)(4,79.85)};
\addlegendentry{q3}
\addplot[color=blue,mark=triangle] coordinates {(1,0.58)(2,0.92)(3,9.06)(4,50.90)};
\addlegendentry{(p3,d4)}
\addplot[color=orange,mark=star] coordinates {(1,24.36)(2,60.97)(3,78.06)(4,81.36)};
\addlegendentry{q4}
\addplot[color=cyan,mark=diamond] coordinates {(1,0.23)(2,0.54)(3,2.83)(4,22.88)};
\addlegendentry{(p4,d3)}
\addplot[color=brown,mark=o] coordinates {(1,3.62)(2,30.77)(3,35.51)(4,53.51)};
\addlegendentry{q5}
\addplot[color=blue!70!black,mark=triangle*] coordinates {(1,0.15)(2,0.32)(3,1.46)(4,13.24)};
\end{axis}
\end{tikzpicture}
\label{fig:vgg-rel}
}
\hfill
\subfloat[ResNet-18]{%
\begin{tikzpicture}
\begin{axis}[
    width=0.32\textwidth,
    height=0.22\textheight,
    xlabel={BER},
    ylabel={Accuracy Drop (\%)},
    ymin=0, ymax=100,
    xtick={1,2,3,4},
    xticklabels={1E-5, 3E-5, 1E-4, 3E-4},
    legend pos=north west,
    ymajorgrids=true,
    grid style=dashed,
    legend style={font=\tiny},
]
\addplot[color=red,mark=square] coordinates {(1,2.79)(2,23.95)(3,64.58)(4,80.23)};
\addlegendentry{q3}
\addplot[color=blue,mark=triangle] coordinates {(1,0.84)(2,3.86)(3,24.3)(4,74.5)};
\addlegendentry{(p3,d4)}
\addplot[color=orange,mark=star] coordinates {(1,1.91)(2,9.82)(3,66.85)(4,79.93)};
\addlegendentry{q4}
\addplot[color=cyan,mark=diamond] coordinates {(1,0.54)(2,2.01)(3,8.25)(4,74.04)};
\addlegendentry{(p4,d3)}
\addplot[color=brown,mark=o] coordinates {(1,7.53)(2,9.10)(3,68.71)(4,78.87)};
\addlegendentry{q5}
\addplot[color=blue!70!black,mark=triangle*] coordinates {(1,0.10)(2,0.88)(3,5.38)(4,66.43)};
\addlegendentry{(p5,d2)}
\end{axis}
\end{tikzpicture}
\label{fig:resnet-rel}
}

\caption{Fault injection results for AlexNet, VGG-11, and ResNet-18 under different BER values.}
\label{fig:combined-rel}
\end{figure*}

The fault injection results depicted in Figures \ref{fig:alex-rel}, \ref{fig:vgg-rel}, and \ref{fig:resnet-rel} show the reliability of the proposed packing and protection scheme across the architectures. For each model, the accuracy degradation is evaluated under varying BERs, ranging from $1\times10^{-5}$ to $3\times10^{-4}$.

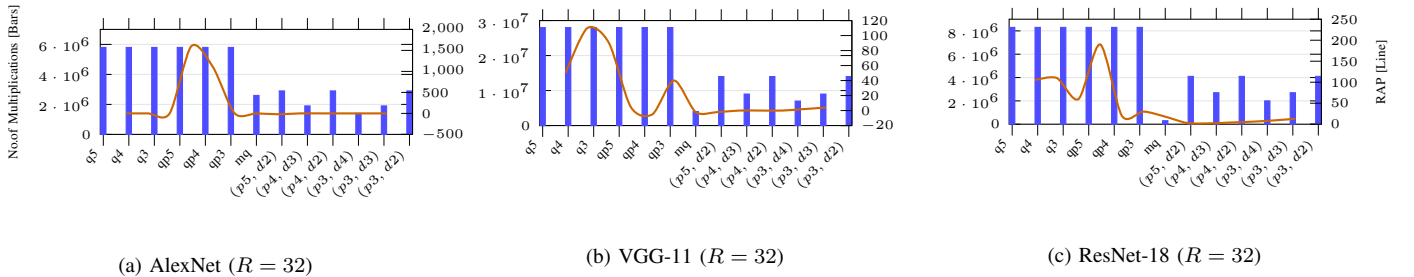
\begin{figure*}[hb]
\centering

\subfloat[AlexNet ($R=32$)]{%
\begin{minipage}{0.315\textwidth}
\centering
\begin{tikzpicture}
\begin{axis}[
    width=\linewidth,   
    height=0.52\linewidth,
    ybar,
    bar width=2pt,
    ymin=0,
    ymax=7000000,
    ylabel={No.of Multiplications [Bars]},
    symbolic x coords={q5,q4,q3,qp5,qp4,qp3,mq,pxyd2,pxyd3,pxyd4,pxyd5,pxyd6,pxyd7},
    xtick=data,
    xticklabels={q5,q4,q3,qp5,qp4,qp3,mq,{$(p5,d2)$},{$(p4,d3)$},{$(p4,d2)$},{$(p3,d4)$},{$(p3,d3)$},{$(p3,d2)$}},
    x tick label style={rotate=45, anchor=east, font=\tiny},
    y tick label style={font=\tiny},
    ylabel style={font=\tiny},
    enlarge x limits=0.01,
    ymajorgrids=true,
    grid style={gray!20},
    axis line style={black},
    tick style={black},
    clip=false,
    scaled y ticks=false,
]
\addplot[fill=blue!70, draw=blue!70] coordinates {
    (q5,5800000)
    (q4,5800000)
    (q3,5800000)
    (qp5,5800000)
    (qp4,5800000)
    (qp3,5800000)
    (mq,2600000)
    (pxyd2,2900000)
    (pxyd3,1900000)
    (pxyd4,2900000)
    (pxyd5,1400000)
    (pxyd6,1900000)
    (pxyd7,2900000)
};
\end{axis}
\begin{axis}[
    width=\linewidth,
    height=0.52\linewidth,
    ymin=-500,
    ymax=2000,
    ytick={-500,0,500,1000,1500,2000},
    axis y line*=right,
    axis x line=none,
    symbolic x coords={q5,q4,q3,qp5,qp4,qp3,mq,pxyd2,pxyd3,pxyd4,pxyd5,pxyd6,pxyd7},
    xtick=data,
    x tick label style={draw=none},
    y tick label style={font=\tiny},
    ylabel style={font=\tiny},
    axis line style={black},
    tick style={black},
    clip=false,
]
\addplot[orange!80!black, thick, smooth, mark=none] coordinates {
    (q5,0)
    (q4,0)
    (q3,0)
    (qp5,1600)
    (qp4,1100)
    (qp3,0)
    (mq,0)
    (pxyd2,-20)
    (pxyd3,0)
    (pxyd4,0)
    (pxyd5,0)
    (pxyd6,0)
    (pxyd7,0)
};
\end{axis}
\end{tikzpicture}
\label{fig:mutl-alex}
\end{minipage}%
}
\hfill
\subfloat[VGG-11 ($R=32$)]{%
\begin{minipage}{0.315\textwidth}
\centering
\begin{tikzpicture}
\begin{axis}[
    width=\linewidth,
    height=0.52\linewidth,
    ybar,
    bar width=2pt,
    ymin=0,
    ymax=30000000,
    symbolic x coords={q5,q4,q3,qp5,qp4,qp3,mq,pxyd2,pxyd3,pxyd4,pxyd5,pxyd6,pxyd7},
    xtick=data,
    xticklabels={q5,q4,q3,qp5,qp4,qp3,mq,{$(p5,d2)$},{$(p4,d3)$},{$(p4,d2)$},{$(p3,d4)$},{$(p3,d3)$},{$(p3,d2)$}},
    x tick label style={rotate=45, anchor=east, font=\tiny},
    y tick label style={font=\tiny},
    ylabel style={font=\tiny},
    enlarge x limits=0.01,
    ymajorgrids=true,
    grid style={gray!20},
    axis line style={black},
    tick style={black},
    clip=false,
    scaled y ticks=false,
]
\addplot[fill=blue!70, draw=blue!70] coordinates {
    (q5,28000000)
    (q4,28000000)
    (q3,28000000)
    (qp5,28000000)
    (qp4,28000000)
    (qp3,28000000)
    (mq,4000000)
    (pxyd2,14000000)
    (pxyd3,9000000)
    (pxyd4,14000000)
    (pxyd5,7000000)
    (pxyd6,9000000)
    (pxyd7,14000000)
};
\end{axis}
\begin{axis}[
    width=\linewidth,
    height=0.52\linewidth,
    ymin=-20,
    ymax=120,
    ytick={-20,0,20,40,60,80,100,120},
    axis y line*=right,
    axis x line=none,
    symbolic x coords={q5,q4,q3,qp5,qp4,qp3,mq,pxyd2,pxyd3,pxyd4,pxyd5,pxyd6,pxyd7},
    xtick=data,
    x tick label style={draw=none},
    y tick label style={font=\tiny},
    ylabel style={font=\tiny},
    axis line style={black},
    tick style={black},
    clip=false,
]
\addplot[orange!80!black, thick, smooth, mark=none, line join=round] coordinates {
    (q5,50)
    (q4,110)
    (q3,90)
    (qp5,5)
    (qp4,-5)
    (qp3,40)
    (mq,-2)
    (pxyd2,-2)
    (pxyd3,0)
    (pxyd4,0)
    (pxyd5,0)
    (pxyd6,2)
    (pxyd7,4)
};
\end{axis}
\end{tikzpicture}
\label{fig:mutl-vgg}
\end{minipage}%
}
\hfill
\subfloat[ResNet-18 ($R=32$)]{%
\begin{minipage}{0.315\textwidth}
\centering
\begin{tikzpicture}
\begin{axis}[
    width=\linewidth,
    height=0.52\linewidth,
    ybar,
    bar width=2pt,
    ymin=0,
    ymax=9000000,
    symbolic x coords={q5,q4,q3,qp5,qp4,qp3,mq,pxyd2,pxyd3,pxyd4,pxyd5,pxyd6,pxyd7},
    xtick=data,
    xticklabels={q5,q4,q3,qp5,qp4,qp3,mq,{$(p5,d2)$},{$(p4,d3)$},{$(p4,d2)$},{$(p3,d4)$},{$(p3,d3)$},{$(p3,d2)$}},
    x tick label style={rotate=45, anchor=east, font=\tiny},
    y tick label style={font=\tiny},
    ylabel style={font=\tiny},
    enlarge x limits=0.01,
    ymajorgrids=true,
    grid style={gray!20},
    axis line style={black},
    tick style={black},
    clip=false,
    scaled y ticks=false,
]
\addplot[fill=blue!70, draw=blue!70] coordinates {
    (q5,8300000)
    (q4,8300000)
    (q3,8300000)
    (qp5,8300000)
    (qp4,8300000)
    (qp3,8300000)
    (mq,300000)
    (pxyd2,4100000)
    (pxyd3,2700000)
    (pxyd4,4100000)
    (pxyd5,2000000)
    (pxyd6,2700000)
    (pxyd7,4100000)
};
\end{axis}
\begin{axis}[
    width=\linewidth,
    height=0.52\linewidth,
    ymin=0,
    ymax=250,
    ytick={0,50,100,150,200,250},
    axis y line*=right,
    axis x line=none,
    ylabel={RAP [Line]},
    symbolic x coords={q5,q4,q3,qp5,qp4,qp3,mq,pxyd2,pxyd3,pxyd4,pxyd5,pxyd6,pxyd7},
    xtick=data,
    x tick label style={draw=none},
    y tick label style={font=\tiny},
    ylabel style={font=\tiny},
    axis line style={black},
    tick style={black},
    clip=false,
]
\addplot[orange!80!black, thick, smooth, mark=none, line join=round] coordinates {
    (q5,105)
    (q4,110)
    (q3,60)
    (qp5,190)
    (qp4,20)
    (qp3,30)
    (mq,18)
    (pxyd2,3)
    (pxyd3,2)
    (pxyd4,4)
    (pxyd5,6)
    (pxyd6,9)
    (pxyd7,13)
};
\end{axis}
\end{tikzpicture}
\label{fig:mutl-resnet}
\end{minipage}%
}
\caption{Number of multiplications (bars, left axis) and RAP values 
(line, right axis) across all evaluated configurations for AlexNet, 
VGG-11, and ResNet-18.}
\label{fig:mutl-combined}
\end{figure*}

In all cases, the unprotected quantized models experience significant accuracy drops as BER increases, particularly beyond $1\times10^{-4}$. The sharp declines in the figure confirm the vulnerability of fully quantized models when no protection is applied, particularly at lower bit-widths. In contrast, the proposed method shows a clear and consistent improvement in fault tolerance. Configurations such as $(p3,d4)$, $(p4,d3)$, and $(p5,d2)$ exhibit significantly smaller accuracy drops across all BERs. In VGG-11, for instance, the $(p3,d4)$ configuration limits the drop to just 9.06\% at BER = $1\times10^{-4}$, compared to 79.80\% in its unprotected counterpart.

Figures \ref{fig:mutl-alex}–\ref{fig:mutl-resnet} further confirm that selective protection with packing reduces multiplications and RAP substantially compared to both unprotected and fully protected baselines.

\subsection{Packing Efficiency Analysis}

To evaluate the execution efficiency gained through Granular SWAR packing, we compare the number of MAC operations required for inference with and without the Safe-FFD packing algorithm. All models are evaluated using a register width of $R = 32$ bits on an FPGA-based systolic array substrate, where $R$ is a configurable design parameter determined by the target hardware. The quantization framework accepts any integer bit-width as input; in this study, configurations at $b \in \{2, 4, 8\}$ bits per operand are evaluated, with sensitivity profiling conducted at each bit-width to construct the optimal bit-width schedule $\mathbf{b}$ for each architecture.

The results demonstrate substantial reductions in MAC operations across all architectures. VGG-11 achieves the highest efficiency gain with 55.9\% MAC reduction, ResNet-18 achieves a 42.9\% while AlexNet shows a 6.5\%.

These efficiency gains arise purely from improved data layout and register utilization, as the PE hardware remains unchanged. The Safe-FFD algorithm packs heterogeneous-precision operand pairs at compile time, enabling Granular SWAR execution where multiple MAC operations are performed per register word. The effective iteration depth $K_{\text{eff}}$ is reduced proportionally to the average packing density $\bar{d}$, directly translating to fewer memory accesses and execution cycles.

\begin{figure}[h]
\centering
\includegraphics[width=0.42\textwidth]{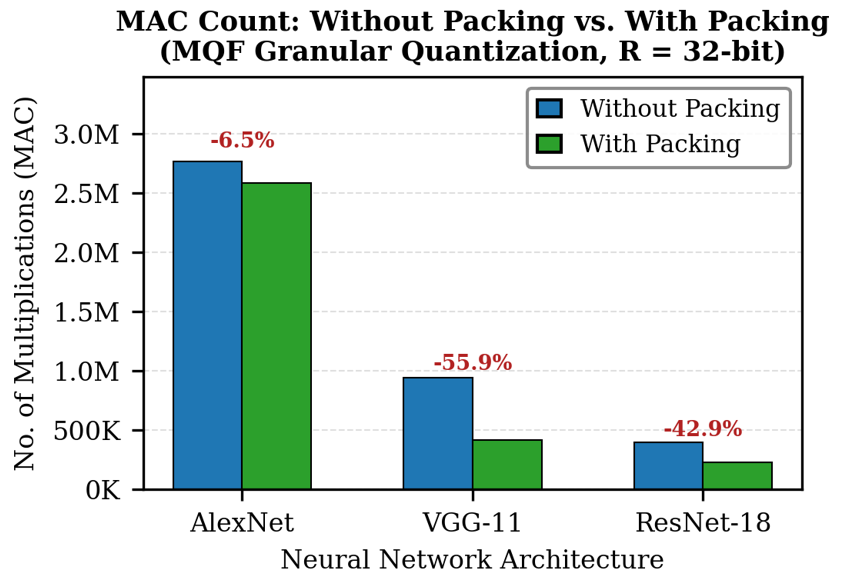}
\caption{MAC count comparison: without packing vs.\ with packing for AlexNet, VGG-11, and ResNet-18.}
\label{fig:mac_comparison}
\end{figure}

Fig.~\ref{fig:mac_comparison} illustrates the MAC count comparison between unpacked and packed execution for the three architectures. The red annotations indicate the percentage reduction achieved through deterministic register packing. Packing efficiency varies across architectures due to differences in layer-wise bit-width distributions produced by sensitivity profiling. VGG-11's higher packing density reflects its more aggressive quantization schedule, where multiple layers are assigned lower bit-widths, enabling tighter packing within the 32-bit register constraint.


\section{Conclusion}
\label{sec:conclusion}

This paper proposes a reliability-aware weight-packing framework that combines sensitivity-driven mixed-precision quantization, deterministic register packing, and selective bit-level protection for DNN acceleration on systolic arrays.

The sensitivity-based profiling method assigns per-layer bit-widths based on accuracy impact without requiring iterative retraining. The Safe-FFD packing algorithm consolidates heterogeneous-precision operand pairs into register words at compile time, enabling Granular SWAR execution that reduces MAC operations by up to 56\% while preserving numerical correctness. Selective MSB protection applies TMR only to vulnerable layers identified through empirical fault injection, achieving fault resilience with minimal overhead.

Experimental evaluations on AlexNet, VGG-11, and ResNet-18 demonstrate up to 62\% memory savings and significant improvements in $RAP$ and $P_{drop}$ metrics compared to baseline approaches. The framework requires no hardware modifications and is fully compatible with FPGA-based systolic array accelerators, enabling efficient deployment of quantized neural networks in resource-constrained, fault-prone edge environments.

\section*{ACKNOWLEDGMENT}
\small
This work was supported in part by the Estonian Research Council grant PUT PRG1467 ``CRASHLESS'', EU Grant Project 101160182 ``TAICHIP'', and by the Federal Ministry of Research, Technology and Space of Germany (BMFTR) for supporting Edge-Cloud AI for DIstributed Sensing and COmputing (AI-DISCO) project (Project-ID ``16ME1127'').

\bibliographystyle{IEEEtran}
\bibliography{ref}

@inproceedings{leveugle2009statistical,
  title={Statistical fault injection: Quantified error and confidence},
  author={Leveugle, R{\'e}gis and et al},
  booktitle={DATE},
  pages={502--506},
  year={2009},
  organization={}
}

@article{2,
  title={Cnn2gate: Toward designing a general framework for implementation of convolutional neural networks on {FPGA}},
  author={Ghaffari, Alireza and Savaria, Yvon},
  journal={arXiv preprint arXiv:2004.04641},
  year={2020}
}

@inproceedings{taheri2024adam0,
  title={{AdAM:} Adaptive fault-tolerant approximate multiplier for edge {DNN} accelerators},
  author={Taheri, Mahdi and others},
  booktitle={2024 IEEE European Test Symposium (ETS)},
  year={2024}
}

@inproceedings{taheri2024exploration,
  title={Exploration of Activation Fault Reliability in Quantized Systolic Array-Based {DNN} Accelerators},
  author={Taheri, Mahdi and others},
  booktitle={2024 25th International Symposium on Quality Electronic Design (ISQED)},
  year={2024}
}

@article{1,
  title={fpgaConvNet: Mapping regular and irregular convolutional neural networks on FPGAs},
  author={Venieris, Stylianos I and Bouganis, Christos-Savvas},
  journal={IEEE transactions on neural networks and learning systems},
  volume={30},
  number={2},
  pages={326--342},
  year={2018},
  publisher={IEEE}
}

@inproceedings{taheri2023deepaxe,
  title={Deepaxe: A framework for exploration of approximation and reliability trade-offs in dnn accelerators},
  author={Taheri, Mahdi and others},
  booktitle={2023 24th International Symposium on Quality Electronic Design (ISQED)},
  pages={1--8},
  year={2023},
  organization={IEEE}
}

@article{nazarifortune,
  title={FORTUNE: A Negative Memory Overhead Hardware-Agnostic Fault TOleRance TechniqUe in DNNs},
  author={Nazari, Samira and others},
  journal={Authorea Preprints},
  year={2024},
  publisher={Authorea}
}

@inproceedings{vlsi-soc,
  title={Heterogeneous Approximation of DNN HW
Accelerators based on Channels Vulnerability},
  author={Cherezova, Natalia and others},
  booktitle={EEE International Conference on Very Large-Scale Integration
(VLSI-SoC)},
  year={2024},
  organization={In press}
}

@inproceedings{jacob2018quantization,
  title={Quantization and training of neural networks for efficient integer-arithmetic-only inference},
  author={Jacob, Benoit and Kligys, Skirmantas and Chen, Bo and Zhu, Menglong and Tang, Matthew and Howard, Andrew and Adam, Hartwig and Kalenichenko, Dmitry},
  booktitle={Proceedings of the IEEE conference on computer vision and pattern recognition},
  pages={2704--2713},
  year={2018}
}

@inproceedings{wang2019haq,
  title={Haq: Hardware-aware automated quantization with mixed precision},
  author={Wang, Kuan and Liu, Zhijian and Lin, Yujun and Lin, Ji and Han, Song},
  booktitle={Proceedings of the IEEE/CVF conference on computer vision and pattern recognition},
  pages={8612--8620},
  year={2019}
}

@inproceedings{sharma2018bit,
  title={Bit fusion: Bit-level dynamically composable architecture for accelerating deep neural network},
  author={Sharma, Hardik and Park, Jongse and Suda, Naveen and Lai, Liangzhen and Chau, Benson and Kim, Joon Kyung and Chandra, Vikas and Esmaeilzadeh, Hadi},
  booktitle={2018 ACM/IEEE 45th Annual International Symposium on Computer Architecture (ISCA)},
  pages={764--775},
  year={2018},
  organization={IEEE}
}

@inproceedings{nazari2025reliability,
  title={Reliability-aware performance optimization of DNN HW accelerators through heterogeneous quantization},
  author={Nazari, Samira and Taheri, Mahdi and Azarpeyvand, Ali and Afsharchi, Mohsen and Herglotz, Christian and Jenihhin, Maksim},
  booktitle={2025 IEEE 26th Latin American Test Symposium (LATS)},
  pages={1--6},
  year={2025},
  organization={IEEE}
}

@inproceedings{nazari2025genie,
  title={Genie: Genetic algorithm-based reliability assessment methodology for deep neural networks},
  author={Nazari, Samira and Taheri, Mahdi and Azarpeyvand, Ali and Afsharchi, Mohsen and Herglotz, Christian and Jenihhin, Maksim},
  booktitle={2025 11th International Conference on Computing and Artificial Intelligence (ICCAI)},
  pages={264--271},
  year={2025},
  organization={IEEE}
}

@inproceedings{won2023ulppack,
  title={ULPPACK: Fast Sub-8-bit Matrix Multiply on Commodity SIMD Hardware},
  author={Won, Jaeyeon and Si, Jeyeon and Son, Sam and Ham, Tae Jun and Lee, Jae W.},
  booktitle={Proceedings of the International Conference on Architectural Support for Programming Languages and Operating Systems (ASPLOS)},
  year={2023}
}

@article{eslami2024mono,
  title={MONO: Enhancing Bit-Flip Resilience With Bit Homogeneity for Neural Networks},
  author={Eslami, Maryam and Liu, Yuhao and Hosseini, Reshad and Ullah, Salim and Mirsalari, Seyed Ahmad and Kumar, Akash and Salehi, Mostafa E.},
  journal={IEEE Transactions on Computer-Aided Design of Integrated Circuits and Systems},
  year={2024},
  publisher={IEEE}
}

@article{fusion,
  title={Fusion: A Framework for Unified Sequential Token AdaptatIon in VisiOn TraNsformers},
  author={Pradeep, Aravind and Nazari, Samira and Taheri, Mahdi and Herglotz, Christian},
  journal={arXiv preprint arXiv:2607.02612},
  year={2026}
}

@incollection{quant,
  title={Quantization-Aided Cost-Efficient Reliability of CNN Accelerators for Edge AI},
  author={Jenihhin, Maksim and Taheri, Mahdi and Cherezova, Natalia},
  booktitle={Machine Learning Systems: The Role of Hardware Design for Dependable Computing},
  pages={255--283},
  year={2026},
  publisher={Springer}
}

\end{document}